\documentclass[review,5p,authoryear,12pt]{elsarticle}

\usepackage{color}
\usepackage[hyphens]{url}
\usepackage[draft]{hyperref}
\usepackage{breakurl}
\usepackage{setspace}

\usepackage{alltt}
\usepackage{float}
\usepackage{graphicx}
\usepackage{epstopdf}

\graphicspath{ {diagrams/}{images/} }
\usepackage{microtype}

\usepackage{soul}

\newcommand{\newtext}[1]{#1}
\newcommand{\oldtext}[1]{}

\journal{Astronomy \& Computing}

\begin{document}

\begin{frontmatter}

\title{Use of Docker for deployment and testing of astronomy software}
\tnotetext[ref:docker]{\burl{https://www.docker.com}}

\author{D.~Morris, S.~Voutsinas, N.C.~Hambly and R.G.~Mann} 
\ead{dmr,stv,nch,rgm@roe.ac.uk}

\address{Institute for Astronomy, School of Physics and Astronomy, University of Edinburgh, Royal Observatory,
Blackford Hill, Edinburgh, EH9~3HJ, UK}

\begin{abstract}
We describe preliminary investigations of using Docker for the deployment and
testing of astronomy software. Docker is a relatively new containerisation
technology that is developing rapidly and being adopted across a range of
domains. It is based upon virtualization at operating system level, which
presents many advantages in comparison to the more traditional hardware
virtualization that underpins most cloud computing infrastructure today. A
particular strength of Docker is its simple format for describing and managing
software containers, which has benefits for software developers, system
administrators and end users.

We report on our experiences from two projects -- a simple activity to
demonstrate how Docker works, and a more elaborate set of services that
demonstrates more of its capabilities and what they can achieve within an
astronomical context -- and include an account of how we solved problems through
interaction with Docker's very active open source development community, which
is currently the key to the most effective use of this rapidly-changing
technology.
\end{abstract}


\end{frontmatter}

\section{Introduction}
\label{sec:intro}

In common with many sciences, survey astronomy has entered the era of
``Big~Data'', which changes the way that sky survey data centres must operate.
For more than a decade, they have been following the mantra of `ship the
results, not the data' \citep[e.g.~][ and other contributions within the same
volume]{2004SPIE.5493..137Q} and deploying ``science archives'' \citep[e.g.~][
and references therein]{2008MNRAS.384..637H}, which provide users with
functionality for filtering sky survey data sets on the server side, to reduce
the volume of data to be downloaded to the users' workstations for further
analysis. Typically these science archives have been implemented in relational
database management systems, and astronomers have become adept in exploiting the
power of their Structured Query Language (SQL) interfaces.

However, as sky survey catalogue databases have grown in size -- the UKIDSS
\citep{2008MNRAS.384..637H} databases were 1--10 terabytes, VISTA 
\citep{2012A&A...548A.119C} catalogue data releases are several 10s of
terabytes, as is the final data release from the Sloan Digital Sky
Survey~(DR12; \citealt{2015ApJS..219...12A}),
Pan-STARRS1 is producing a $\sim$100 terabyte
database \citep{2015IAUGA..2258174F} and LSST~(\citealt{2015arXiv151207914J};
catalogue data volumes of up to 1 terabyte per night) will produce
databases several petabytes in size -- the minimally useful subset of data for users is
growing to the point where simple filtering with an SQL query is not
sufficient to generate a result set of modest enough size for a user to want to
download to their workstation. This means that the data centre must provide the
server--side computational infrastructure to allow users to conduct (at least
the first steps in) their analysis in the data centre before downloading
a small result set. The same requirement arises for data centres that wish to
support survey teams in processing their imaging data (with data volumes
typically~10 to~20 times larger than those quoted above for catalogue data sets).
In both cases data centre staff face practical issues when supporting different
sets of users running different sets of software on the same physical infrastructure
(e.g.~\citealt{2009ASPC..411..185G}).

These requirements are not, of course, peculiar to astronomy, and similar
considerations have motivated the development of Grid and Cloud Computing
over the past two decades. A pioneering example of the deployment of cloud
computing techniques for astronomy has been the CANFAR project
\citep{2009ASPC..411..185G, 2011ASPC..442...61G, 2015scop.confE..38G} undertaken
by the Canadian Astronomy Data Centre and collaborators in the Canadian research
computing community. The current CANFAR system is based on hardware
virtualization, where the data processing software and web services
are run in virtual machines, isolated from the details of the underlying
physical hardware.

Following on from the development of systems based on hardware virtualization,
the past few years have seen an explosion of interest within both the research
computing and commercial IT sectors in operating system level virtualization,
which provides an alternative method of creating and managing the virtualized
systems.

A lot of the most recent activity in this field has centred on Docker and this
paper presents lessons learned from two projects we have conducted involving
Docker: a simple test of its capabilities as a deployment system and as part of
more complicated project connecting a range of Virtual Observatory~(VO;
\citealt{ivoaarch}) services running in separate Docker containers.

Even by the standards of large open source projects, the rise of Docker has been
rapid and its development continues apace. A journal paper cannot hope,
therefore, to present an up-to-date summary of Docker, nor an authoritative
tutorial in its use, so we attempt neither here. Rather, we aim to describe the
basic principles underlying Docker and to contrast its capabilities with the
virtual machine technologies with which astronomers may be more familiar,
highlighting where operating system level virtualization provides benefit for
astronomy. We illustrate these benefits through describing our two projects and
the lessons we have learned from undertaking them. Many of the issues we
encountered have since been solved as the Docker engine and toolset continue to
evolve, but we believe there remains virtue in recounting them, both because
they illustrate basic properties of Docker containers and because they show how
the Docker community operates.

For the sake of definiteness, we note the development of the systems described
in this paper were based on Docker version 1.6 and that we discuss solutions to
the issues we encountered that have appeared up to version 1.12.

The plan of this paper is as follows. Section~\ref{sec:vmcont} describes
hardware and operating system level virtualization, summarising the
differences between the two approaches. Section~\ref{sec:docker} introduces
Docker as a specific implementation of operating system level virtualization.
Section~\ref{sec:ivoatex} describes our first Docker project, in which it was
used to create a deployment system for the IVOA\TeX \ Document Preparation System
\citep{ivoatex}. Section~\ref{sec:firethorn} describes the use of Docker in the
development and deployment of the Firethorn VO data access service
\citep{morris_2013}. Section~\ref{sec:issues} describes a specific problem we
encountered and how it was solved through interaction with the Docker
development community. Section~\ref{sec:conclusion} summarises some of the
lessons we learned from these experiments and presents our conclusions as to the
place that Docker (or similar technologies) may develop in astronomical data
management.

\section{Virtual machines and containers}
\label{sec:vmcont}

The physical hardware of a large server may have multiple central processor
units, each with multiple cores with support for multiple threads and access
to several hundred gigabytes of system memory. The physical hardware may also
include multiple hard disks in different configurations, including software
and hardware RAID systems, and access to network attached storage.
However, it is rare for a software application to require direct access to
the hardware at this level of detail.
In fact, it is more often the case that a software application's hardware
requirements can be described in much simpler terms.
Some specific cases such as database services dealing with large data sets
may have specific hardware requirements for disk access but in most cases this
still would represent a subset of the hardware available to the physical machine.

Virtualization allows a system administrator to create a virtual environment for
a software application that provides a simplified abstract view of the
system. If a software application is able to work within this abstract environment
then the same application can be moved or redeployed on any platform
that is capable of providing the same virtual environment, irrespective of what
features or facilities the underlying physical hardware provides.
This ability to create standardized virtual systems on top of a variety of
different physical hardware platforms formed the basis of the Infrastructure as
a Service (IaaS) cloud computing service model as exemplified by the large scale
providers like Amazon Web Services (AWS)\footnote{\burl{https://aws.amazon.com/}}.
The interface between customer and service provider is based on provision of
abstract virtual machines. The details of the underlying hardware platform and
the infrastructure required to provide network, power and cooling are all the
service provider's problem.
What happens inside the virtual machine is up to the customer, including
the choice of operating system and software applications deployed in it.

With hardware virtualization, each virtual machine includes a simulation
of the whole computer, including the system BIOS, PCI bus, hard disks, network
cards, etc. The aim is to create a detailed enough simulation such that the
operating system running inside the virtual machine is not aware that it is running in a
simulated environment.
The key advantage of this approach is that because the guest system is isolated
from the host, the guest virtual machine can run a completely different
operating system to that running on the physical host.
However, this isolation comes at a price. With hardware virtualization
each virtual machine uses a non--trivial share of the physical system's resources just
implementing the simulated hardware, resources which are no longer available for
running the intended application software and services.
Most of the time this cost is hidden from the end user, but it is most visible
when starting up a new virtual machine. With hardware virtualization the virtual machine has to
run through the full startup sequence from the initial BIOS boot through to the
guest operating system initalization process, starting the full set of daemons
and services that run in the background.

\begin{figure*}
\includegraphics[width=\textwidth]{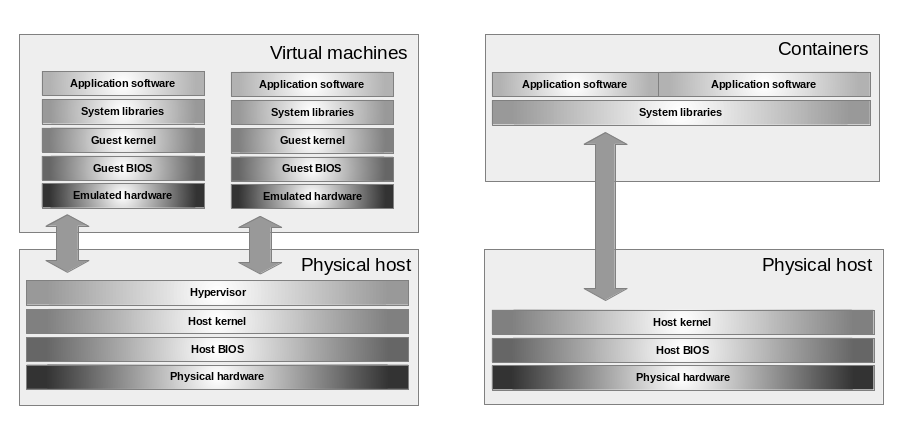}
\caption{Comparison between hardware (left) and operating system level
virtualization (right) .\label{fig:vmcompare}}
\end{figure*}

Comparing hardware virtualization with operating system level
virtualization (Figure~\ref{fig:vmcompare}) we find
a number of key differences between them, to do with what they are capable of
and how they are used. A key difference is determined by the different technologies used to
implement the virtual machines.
In hardware virtualization the host
system creates a full simulation of the guest system, including the system
hardware, firmware and operating system.
With operating system level virtualization the physical host operating system
and everything below it, including the hardware and system firmware, is
shared between the host and guest systems.
This imposes a key limitation on operating system level virtualization in that
the host and guest system must use the same operating system. So, for example, if a Linux
host system can use operating system level virtualization to support guests running
different Linux distributions and versions, it cannot use operating system level
virtualization to support a Berkeley Software Distribution (BSD) or Illumos guest. 
However, if this limitation is not a problem, then sharing the system hardware,
firmware and operating system kernel with the host system means that supporting
operating system level virtual machines, commonly referred to as {\em
containerisation}, represents a much lower cost in terms of system resources.
This, in turn leaves more of the system resources available for running the
intended application software and services.

\section{Docker}
\label{sec:docker}

Docker is emerging as the technology of choice for~VM
containers~\citep{2015arXiv150908231Y,2015ISPAn.II4...43W}. 
Docker is an operating system level virtualization environment
that uses software containers to provide isolation between applications.
The rapid adoption and evolution of Docker from the initial open
source project launched in 2013\footnote{\burl{http://www.infoq.com/news/2013/03/Docker}}
by `platform--as--a--service' (PaaS) provider dotCloud\footnote{\burl{https://www.dotcloud.com/}}, to the
formation of the Open Container Initiative
(OCI)\footnote{\burl{https://www.opencontainers.org/}} in
2015\footnote{\burl{http://blog.docker.com/2015/06/open-container-project-foundation/}} suggests that Docker met a real need within the software development community
which was not being addressed by the existing tools. (As an aside, it is interesting
to note that the technologies behind OS virtualization have been available for a
number of years. For example,
Solaris containers have been available as part of the Solaris operating system
since 2005, and {\em cgroups} and {\em namespaces} have been part of the Linux
kernel since 2007.)

Although both the speed and simplicity of Docker containers have been factors
contributing to its rapid adoption, arguably it is the development of a
standardized \texttt{Dockerfile} format for describing and managing software
containers that has been the unique selling point, differentiating Docker from
its competitors\footnote{\burl{http://www.zdnet.com/article/what-is-docker-and-why-is-it-so-darn-popular/}}\textsuperscript{,}\footnote{\burl{http://www.americanbanker.com/news/bank-technology/why-tech-savvy-banks-are-gung-ho-about-container-software-1078145-1.html/}},
and has been the main driving force behind the rapid adoption of Docker across
such a wide range of different applications:

\begin{itemize}

  \item At the end user level, Docker enables users to describe, share and
  manage applications and services using a common interface by wrapping them in
  standardized containers.

  \item From a developer's perspective, Docker makes it easy to create standard
  containers for their software applications or services.

  \item From a system administrator's perspective, Docker makes it easy to
  automate the deployment and management of business level services as a
  collection of standard containers.

\end{itemize}

\subsection{Docker, DevOps and MicroServices}  

In a `DevOps' (development and operations) environment, software developers and
system administrators work together to develop, deploy and manage enterprise
level services. Describing the deployment environment using Docker containers
enables the development team to treat system infrastructure as code, applying
the same tools they use for managing the software source code, e.g.~source
control, automated testing, etc.~to the infrastructure configuration. 

\subsection{Reproducible science}  

In the science and research community, Docker's ability to describe a software
deployment environment has the potential to improve the reproducibility and the
sharing of data analysis methods and techniques:

\begin{itemize}

  \item \cite{2014arXiv1410.0846B} describes how the ability to publish a
  detailed description of a software environment alongside a research paper
  enables other researchers to reproduce and build on the original work.

  \item \cite{2015arXiv150908789N} describes work to develop containerized
  versions of software tools used to analyse data from particle
  accelerators\footnote{\burl{https://github.com/radiasoft/containers}}.

  \item The Nucletid project\footnote{\burl{http://nucleotid.es/}} provides
  reproducible evaluation of genome assemblers using docker containers.

  \item The BioDocker\footnote{\burl{http://biodocker.org/docs/}} project
  provides a curated set of bioinformatics software using Docker containers.

\end{itemize}

\subsection{Compute resource services}  

There are two roles in which Docker may be useful in implementing systems which
enable end users to submit their own code to a compute resource for execution
within a data centre.
Docker can be used internally as part of the virtualization layer for deploying
and managing the execution environments for the submitted code.
This scenario is already being evaluated by a number of groups, in particular
Docker is one of the technologies being used to deliver a PaaS infrastructure 
for the European Space Agency's Gaia mission archive \citep{O1-4_adassxxv,P080_adassxxv}.

Alternatively, Docker can be used as part of the public service interface,
providing the standard language for describing and packaging the software.
In this scenario, the user would package their software in a container and then
either submit the textual description or the binary container image to the
service for execution.
The advantage of this approach is that the wrapping of analysis software in a
standard container enables the user to build and test their software on their
own platform before submitting it to the remote service for execution.
The common standard for the container runtime environment means that the user
can be confident that their software will behave in the same manner when tested
on a local platform or deployed on the remote service.
\newtext{For a service provider, using Docker to add containerization to the
virtualization layer makes it easier to provide reliable, predictable
execution environments for users to deploy their code into. This in turn reduces
the number of support issues regarding software deployment and installation that
the service provider needs to deal with.}

\subsection{Reproducible deployment} 

It is often the case that a development team does not have direct control over
the software environment where their service will be deployed.
For example, the deployment platform may be configured with versions
of operating system, Java runtime and Tomcat webserver which are determined by the
requirements of other applications already running on the machine and by the
system administrators running the system.
This can present problems when attempting to update the version of these
infrastructure components. Unless it is possible to isolate the different
components from each other then a system component cannot be updated unless all
of the other components that interact with it can be updated at the same time.

With an operating system level virtualization technology like Docker, each
application can be wrapped in a container configured with a specific version of operating system,
language runtime or webserver.
The common interface with the system is set at the container level, not at the
operating system, language or application server level.
In theory it is possible to do something similar using hardware
virtualization. However, in practice the size and complexity
of the virtual machine image makes this difficult.

In a container based approach to service deployment, the development process
includes a container specifically designed for the service.
The same container is used during the development and testing of the software
and becomes part of the final project deliverable.
The final product is shipped and deployed as the container, with all of its
dependencies already installed, rather than as an individual software component
which requires a set of libraries and components that need to be installed
along with it.
This not only simplifies the deployment of the final product, it also makes it
more reproducible.

\section{IVOA\TeX\ in Docker}
\label{sec:ivoatex}

As an early experiment in using containers to deploy applications, we used
Docker to wrap the IVOA\TeX\footnote{\burl{http://www.ivoa.net/documents/Notes/IVOATex}}
document build system to make it easier to use.
The IVOA\TeX\ system uses a combination of \LaTeX\ tools and libraries, a
compiled C program to handle \LaTeX\ to HTML conversion, and a \texttt{makefile}
to manage the build process.

IVOA\TeX\ includes a fairly clear set of install instructions. However,
the instructions are specific to the Debian Linux distribution and porting them
to a different Linux distribution is not straightforward.
In addition, it was found that, in some instances, configuring a system with the
libraries required by the IVOA\TeX\ system conflicted with those required by
other document styles.   

Installing the full IVOA\TeX\ software makes sense for someone who would be
using it regularly. However, installing and configuring all of the required
components is a complicated process for someone who just wants to make a small
edit to an existing IVOA document.
In order to address this we created a simple Docker container that incorporates all
of the components needed to run the IVOA\TeX\ system configured and ready to
run.

The source code for our IVOA\TeX\ container is available on
GitHub\footnote{\burl{https://github.com/ivoa/ivoatex}} and a binary image is
available from the Docker
registry\footnote{\burl{https://hub.docker.com/r/ivoa/ivoatex/}}.

\newtext{
The Docker Hub\footnote{\burl{https://hub.docker.com}} is a service
provided by Docker Inc to enable users to publish binary images of their containers.
}

\newtext{
The Docker Hub contains curated images for a wide range of
different software including Linux distributions like Debian and Fedora,
programming languages like Java and Python, and database services like
MariaDB and PostgreSQL  alongside user contributed images like our own
IVOA\TeX\ container.
}

\newtext{
In our experience, the wide range of free to use, open source
software available in the Docker Hub made it extremely easy to get started,
simply by running one of these pre-configured images or by using them as the
starting point for developing our own images.
}

The source code for our IVOA\TeX\ project consists of a text \texttt{Dockerfile}
which lists the steps required to create the binary image.

\newtext{
A \texttt{Dockerfile} specifies a list of commands needed to create an
image, thus defining the `recipe' of an image, which is machine-readable, but
still simple enough to be human readable.
}

\newtext{
Although the idea of an automated software deployment definition is not
new, and similar ideas have been developed before, for example the
\texttt{kickstart} format which is part of the Anaconda install tools for RedHat
Linux\footnote{\burl{http://fedoraproject.org/wiki/Anaconda/Kickstart}}.
The \texttt{Dockerfile} is one of the few formats that is simple enough to be
able to describe the configuration for a range of different Linux
distributions using the same basic syntax.
}

\newtext{When a \texttt{Dockerfile} is built, the resulting binary image
consists of a series of layers. Each layer in the image describes the files
which have been added, changed or deleted by the corresponding
\texttt{Dockerfile} command.}

\newtext{By formatting the binary image as a series of layers, each of which has
a unique identifier, the Docker system can share layers that are common between
different containers. So, for example, if two containers are based on the same
parent container, then the Docker host only has to download and store one copy
of the layers needed to build the parent container, and then apply the specific
changes needed to create each of the child images.}

\newtext{The following section describes the \texttt{Dockerfile} commands used
to create the IVOA\TeX \ container.} ({\em N.B.\/} in this and subsequent
listings we have added line numbers to aid explanation; they are not present in
the \texttt{Dockerfile} itself.)

\begin{singlespacing}
\begin{alltt}
 1   # Set our base image and maintainer.
 2   FROM metagrid/notroot-debian
 3   MAINTAINER <user@example.com>
 4 
 6   # Disable interactive install.
 7   ENV DEBIAN_FRONTEND noninteractive
 8   
 9   # Install our C build tools.
10   RUN apt-get update \textbackslash
11     && apt-get -yq install \textbackslash
12     zip \textbackslash
13     make \textbackslash
14     gcc \textbackslash
15     libc-dev
16
17   # Install our HTML tools.
18   RUN apt-get update \textbackslash
19     && apt-get -yq install \textbackslash
20     xsltproc \textbackslash
21     libxml2-utils \textbackslash
22     imagemagick \textbackslash
23     ghostscript
24
25   # Install our LaTex tools.
26   RUN apt-get update \textbackslash
27     && apt-get -yq install \textbackslash
28     texlive-latex-base \textbackslash
29     texlive-latex-extra \textbackslash
30     texlive-bibtex-extra \textbackslash
31     texlive-fonts-recommended \textbackslash
32     latex-xcolor \textbackslash
33     cm-super
34
35   # Set our username and home directory.
36   ENV username texuser
37   ENV userhome /var/local/texdata
38
39   # Make our home directory a volume.
40   VOLUME /var/local/texdata
\end{alltt}
\end{singlespacing}


The first line of a \texttt{Dockerfile} uses the \texttt{FROM} command to
declare the parent image this project is derived from:

\begin{singlespacing}
\begin{alltt}
 1   # Set our base image and maintainer
 2   FROM metagrid/notroot-debian
\end{alltt}
\end{singlespacing}

At the start of the build process, Docker will download this base from the
Docker registry and then apply our build instructions to it.
Each new instruction adds another layer in the file system of the final
image.

In our example, the \texttt{notroot-debian} image is a container
developed by one of our colleagues that includes tools for changing the user
account when the container is started. This enables our \LaTeX\ tools
to run using our normal user account rather than root.

The next section of the \texttt{Dockerfile} uses the \texttt{ENV} command to set
an environment variable that prevents the \texttt{apt-get} install commands
from requesting interactive user input:

\begin{singlespacing}
\begin{alltt}
 6   # Disable interactive install.
 7   ENV DEBIAN_FRONTEND noninteractive
\end{alltt}
\end{singlespacing}

A \texttt{Dockerfile} may contain multiple \texttt{ENV} commands to set
environment variables which will be available both during the build process
and in the runtime environment for a container. 
 
The next section uses \texttt{RUN} commands to call the Debian package manager,
\texttt{apt-get}, to install the C compiler and build tools:

\begin{singlespacing}
\begin{alltt}
 9   # Install our C build tools.
10   RUN apt-get update \textbackslash
11     && apt-get -yq install \textbackslash
12     zip \textbackslash
13     make \textbackslash
14     gcc \textbackslash
15     libc-dev
\end{alltt}
\end{singlespacing}

Followed by the HTML editing tools:

\begin{singlespacing}
\begin{alltt}
17   # Install our HTML tools.
18   RUN apt-get update \textbackslash
19     && apt-get -yq install \textbackslash
20     xsltproc \textbackslash
21     libxml2-utils \textbackslash
22     imagemagick \textbackslash
23     ghostscript
\end{alltt}
\end{singlespacing}

and the \LaTeX tools:

\begin{singlespacing}
\begin{alltt}
25   # Install our LaTex tools.
26   RUN apt-get update \textbackslash
27     && apt-get -yq install \textbackslash
28     texlive-latex-base \textbackslash
29     texlive-latex-extra \textbackslash
30     texlive-bibtex-extra \textbackslash
31     texlive-fonts-recommended \textbackslash
32     latex-xcolor \textbackslash
33     cm-super
\end{alltt}
\end{singlespacing}

The next two lines use \texttt{ENV} commands to set the default
user name and the home directory which are used by the
\texttt{metagrid/notroot-debian} base image to set the user account
and home directory when a new container is created:

\begin{singlespacing}
\begin{alltt}
35   # Set our username and home directory.
36   ENV username texuser
37   ENV userhome /var/local/texdata
\end{alltt}
\end{singlespacing}

The final step in the build instructions declare the
working directory as a data volume:

\begin{singlespacing}
\begin{alltt}
39   # Make our home directory a volume.
40   VOLUME /var/local/texdata
\end{alltt}
\end{singlespacing}

This marks the \texttt{/texdata} directory as a separate volume outside the
layered file system of the Docker image.
When we run an instance of this container, we can use the \texttt{--volume}
option to mount a directory from the host system as the \texttt{/texdata}
directory inside the container:

\begin{singlespacing}
\begin{alltt}
    docker run \textbackslash
        --volume "\$(pwd):/texdata" \textbackslash 
        "ivoa/ivoatex"
\end{alltt}
\end{singlespacing}

and once inside the container we can use the \texttt{make} commands to
build our IVOA\TeX\ document:

\begin{singlespacing}
\begin{alltt}
    cd /texdata
      make clean
      make biblio
      make
\end{alltt}
\end{singlespacing}

The initial idea for this project was based on the work by Jessie
Frazelle on wrapping desktop applications in
containers\footnote{\burl{https://blog.jessfraz.com/post/docker-containers-on-the-desktop/}}.

While exploring this technique we encountered a security issue that potentially
allows priviledged access to the host file system.

When run from the command line, the Docker \texttt{run} command does not run the
container directly, instead it uses a socket connection to send the command
to the Docker service, which runs the container on your behalf.
A side effect of the way that the Docker service works is that the \emph{root} 
user inside the container also has \emph{root} privileges when accessing the
file system outside the container.
This normally is not a problem, unless you use a \texttt{--volume} option to
make a directory on the host platform accessible from inside the container,
which is exactly what we need to do to enable the IVOA\TeX \ tools to access
the source for our \LaTeX\ document.

In our case, this does not prevent our program from working, but it does mean
that the resulting PDF and HTML documents end up being owned by \emph{root},
which make it difficult for the normal user to manage them.

This is where the user management tools provided by the \texttt{notroot-debian}
base image can help. The source code for the
\texttt{notroot-debian}\footnote{\burl{https://github.com/metagrid/notroot}}
consists of a \texttt{Dockerfile} which describes how to build the image, plus a
shell script that is run when a container instance starts up:

\begin{singlespacing}
\begin{alltt}
 1   FROM debian:wheezy
 2   MAINTAINER <user@example.com>
 3
 4   # Disable interactive install.
 5   ENV DEBIAN_FRONTEND noninteractive
 6
 7   # Install sudo.
 8   RUN apt-get update \textbackslash
 9     && apt-get -yq install \textbackslash
10     sudo
11
12   # Install the notroot script.
13   COPY notroot.sh /notroot.sh
14   RUN  chmod a+x,a-w /notroot.sh
15
16   # Set notroot as the entrypoint.
17   ENTRYPOINT ["/notroot.sh"]
\end{alltt}
\end{singlespacing}
 
As with our \texttt{ivoatex} container, the first line of the
\texttt{Dockerfile} uses the \texttt{FROM} command to declare the base image
to use as the starting point to build the new image. In this case, it
refers to one of the official Debian images registered in the Docker registry: 

\begin{singlespacing}
\begin{alltt}
 1   FROM debian:wheezy
\end{alltt}
\end{singlespacing}

Followed by an \texttt{ENV} and \texttt{RUN} command to install the
\texttt{sudo} program:

\begin{singlespacing}
\begin{alltt}
 4   # Disable interactive install.
 5   ENV DEBIAN_FRONTEND noninteractive
 6
 7   # Install sudo.
 8   RUN apt-get update \textbackslash
 9     && apt-get -yq install \textbackslash
10     sudo
\end{alltt}
\end{singlespacing}

In this example we are installing a tool that most people would normally expect
to be installed by default as part of a normal Debian system.
Many of the base images provided in the Docker registry contain the
minimum set of components necessary to run a basic shell and very little else.
This is by design, both to keep the physical size of the image as small as
possible (85M bytes for the Debian Wheezy base image), and to minimize the
potential attack surface of software that is not required.  
In general it is easier to start with a minimal configuration and add
the components that you need, rather than starting from a larger base and
removing the ones that you do not.

The next section of the \texttt{Dockerfile} uses the \texttt{COPY} command to
copy the shell script into the container image and then uses a \texttt{RUN}
command to set the permissions to make it executable:

\begin{singlespacing}
\begin{alltt}
12   # Install the notroot script.
13   COPY notroot.sh /notroot.sh
14   RUN  chmod a+x,a-w /notroot.sh
\end{alltt}
\end{singlespacing}

The last line of the \texttt{Dockerfile} adds the shell script as the container
\texttt{ENTRYPOINT}. Which means that this script will be invoked whenever a new
container instance is started:

\begin{singlespacing}
\begin{alltt}
16   # Set notroot as the entrypoint.
17   ENTRYPOINT ["/notroot.sh"]
\end{alltt}
\end{singlespacing}

The script itself checks to see if the target user account and group are already
defined, and if not it will create a new user account or group. It then uses 
\texttt{sudo} to change from the \texttt{root} user to the target
user before executing the original command for the container.

If a new \texttt{ivoatex}\ container is run using the following command:  

\begin{singlespacing}
\begin{alltt}
    docker run 
        --env "useruid=\$(id -u)" \textbackslash
        --volume "\$(pwd):/texdata"\textbackslash 
        "ivoa/ivoatex"
\end{alltt}
\end{singlespacing}

The \texttt{--env} option sets the \texttt{useruid} environment variable
to the same \texttt{uid} as the current user.
The \texttt{ENTRYPOINT} script from the \texttt{notroot-debian} image will
use this to create a new user account inside the container with the same
\texttt{uid} as the user outside the container.

The \texttt{--volume} option mounts the current working directory, returned
by the \texttt{pwd} command, as \texttt{/texdata} inside the container.

The result is that the IVOA\TeX\ tools are run inside the container
using the same \texttt{uid} as the external user, and can see the \LaTeX\
document source in the \texttt{/texdata} directory inside the container.

This workaround highlights a potentially serious problem with the way the Docker
system operates.

If we create a standard Debian container and mount the \texttt{/etc} directory
from the host system as \texttt{/albert} inside the container:

\begin{singlespacing}
\begin{alltt}
    docker run \textbackslash
        --volume "/etc:/albert" \textbackslash 
        "debian" \textbackslash
            bash
\end{alltt}
\end{singlespacing}

Then, inside the container we run the \texttt{vi} text editor and edit the
\texttt{passwd} file in the \texttt{/albert} directory:

\begin{singlespacing}
\begin{alltt}
    vi /albert/passwd
\end{alltt}
\end{singlespacing}

The \texttt{--volume} mount means that \texttt{vi} running inside
the container is able to edit the \texttt{passwd} file outside the container,
using \emph{root} privileges from inside the container.

It is important to note that this issue is not caused by a security weakness in
the container or in the Docker service.
The problem occurs because the user that runs a container has direct control
over what resources on the host system the container is allowed to access.
Without the \texttt{--volume} mount, the container would not be able to access
any files on the host system and there would be no problem.
This is not normally an issue, because users would not normally have sufficient
privileges to run Docker containers from the command line.

Users on a production system would normally be given access to a container
management program such as
Kubernetes\footnote{\burl{http://kubernetes.io/}}
or
OpenStack\footnote{\burl{https://www.openstack.org/}}
to manage their containers.
In addition, most Linux distributions now have security constraints in place
which prevent containers from accessing sensitive locations on the file system.
For example, on RedHat based systems the SELinux security module prevents
containers from accessing a location on the file system unless it has explicitly
been granted permission to do so.

\newtext{Developing the IVOATEX container was an experimental project to learn
how Docker works. The privileged escalation issue we encountered relates
to a specific use case, where the end user is launching a user application
container directly from the command line.}

Since we first worked on this, container technology has continued to
evolve and there has been significant progress in a number of areas that
addresses this issue.
In particular the work within Docker on user 
namespaces\footnote{\burl{https://integratedcode.us/2015/10/13/user-namespaces-have-arrived-in-docker/}}\textsuperscript{,}\footnote{\burl{https://docs.docker.com/engine/reference/commandline/dockerd/\#daemon-user-namespace-options}},
but also the work in the Open Containers project
\footnote{\burl{https://runc.io/}}
\footnote{\burl{https://github.com/opencontainers/runc/issues/38}}\textsuperscript{,}\footnote{\burl{https://blog.jessfraz.com/post/getting-towards-real-sandbox-containers/}},
\newtext{
and new container hosting
platforms such as Singularity\footnote{\burl{http://singularity.lbl.gov/}}
which enable users to run Docker containers as non privileged users.
}

\newtext{
If we were to develop a similar user application container in the
future we would probably use a platform like Singularity to run the
container as a non-privileged user, thus avoiding the issue of privileged access
to the file system.
}

\section{Docker in Firethorn}
\label{sec:firethorn}

\subsection{Firethorn overview}

The goal of the Firethorn project is to enable users to run queries and store
results from local astronomy archive or remote IVOA relational databases and
share these results with others by publishing them via a TAP
service\footnote{\burl{http://www.ivoa.net/documents/TAP/}}. 
The project has its origins in a prototype data resource federation 
service \citep{2012ASPC..461..359H} and is built around the Open Grid Service
Architecture Data Access Infrastructure (OGSA-DAI; \citealt{2011ASPC..442..579H}
and references therein).

\begin{figure*}
\includegraphics[width=\textwidth,clip=true,trim=5 0 0 5]{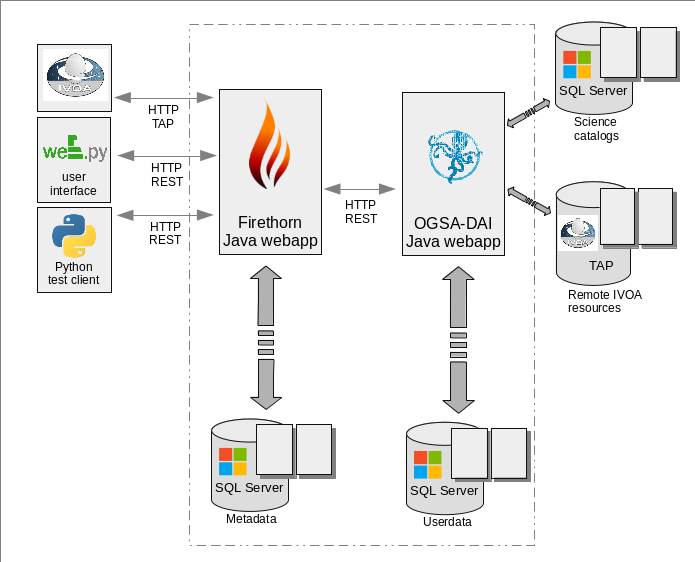}
\caption{Firethorn architecture illustrating the key components and web services.\label{fig:firethorn}}
\end{figure*}

The system architecture consists of two Java web services, one for
handling the catalog metadata, and one for handling database 
queries and processing the results; two SQL Server
databases\footnote{\burl{https://www.microsoft.com/en-us/sql-server/sql-server-2016}},
one for storing the catalog metadata and one for storing query results; a
web.py\footnote{\burl{http://webpy.org/}} based user interface, and a Python
testing tool. A schematic representation of the architecture is shown in
Figure~\ref{fig:firethorn}.

\subsection{Virtual machine allocation and Containerization}

\newtext{
During the development of the Firethorn project we went through a number of
stages in our use of virtualization and containerization. From the initial
development where the services were manually deployed to an automated
system using shell scripts to manage multiple deployments on the same platform:
}
\newtext{
\begin{itemize}
	\item Manually configured virtual machines.
	\item Shell scripts to manage the virtual machines.
	\item Containerization for the core Tomcat web services.
	\item Ambassador pattern for database connections.
	\item Containerizing the Python GUI webapp and the Python test tools.
	\item Orchestration scripts to manage multiple deployments on the same
	platform.
\end{itemize}
}

\begin{figure}[H]
\includegraphics[width=\columnwidth,clip=true,trim=5 0 0 5]{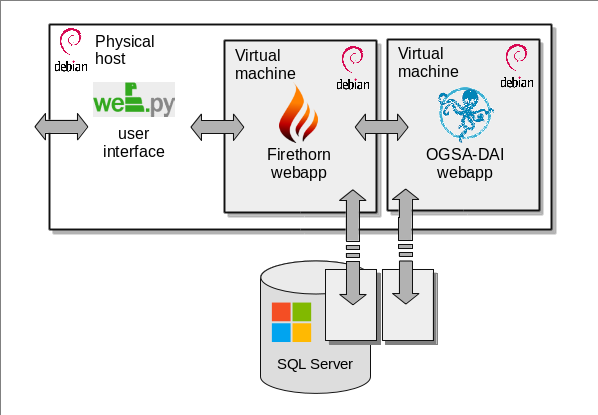}
\caption{Manually configured virtual machine for each web application.
\label{fig:vmmanualconfig}}
\end{figure}

At the beginning of the project we assigned a full
KVM\footnote{\burl{http://www.linux-kvm.org/page/Main_Page}}
virtual machine to each of our Java web services, connected to a Python webapp
running on the physical host which provided the user interface web pages (see
Figure~\ref{fig:vmmanualconfig}; each virtual machine was manually configured).

Assigning a full virtual machine to each component represented a fairly heavy
cost in terms of resources. However, at the time, this level of isolation was
needed to support the different versions of Python, Java and Tomcat required by
each of the components. Using virtual machines like this gave us an initial
level of isolation from the physical host machine configuration. In theory it
also allowed us to run more than one set of services on the same physical
platform, while still being able to configure each set of services independently
without impacting other services running on the same physical hardware.

\begin{figure}[H]
\includegraphics[width=\columnwidth,clip=true,trim=5 0 0 5]{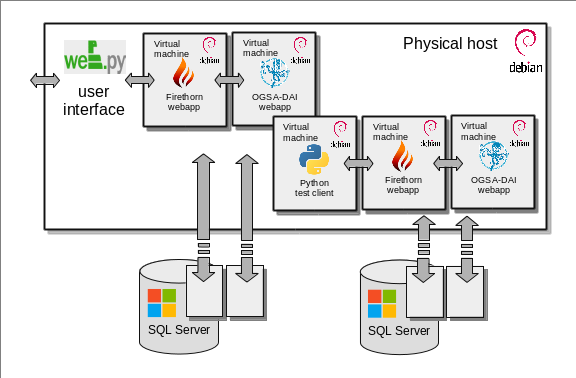}
\caption{Multiple sets of scripted virtual machine configurations.
\label{fig:vmscriptconfig}}
\end{figure}

However, in practice it was not until we moved from using manually configured
virtual machines to using a set of shell scripts based on the
ischnura-kvm\footnote{\burl{https://github.com/Zarquan/ischnura-kvm}}
project to automate the provisioning of new virtual machines that we were able
to run multiple sets of services in parallel. Replacing the manually configured
instances with template based instances gave us the reliable and consistent
set of platforms we needed to develop our automated integration tests (see
Figure~\ref{fig:vmscriptconfig}).

The ischnura-kvm templates handle the basic virtual machine configuration
such as cpu and memory allocation, network configuration, disk space and
operating system.

Once the virtual machines were created, we used a set of shell scripts to
automate the installation of the software packages needed to run each of our
services.
For our Java web services, this included installing and configuring specific
versions of the Java runtime\footnote{\burl{http://openjdk.java.net/}}
and Apache Tomcat\footnote{\burl{http://tomcat.apache.org/}}.
The final step in the process was to deploy our web service and configure them with
the user accounts and passwords needed to access the local databases.

\begin{figure}[H]
\includegraphics[width=\columnwidth,clip=true,trim=5 0 0 5]{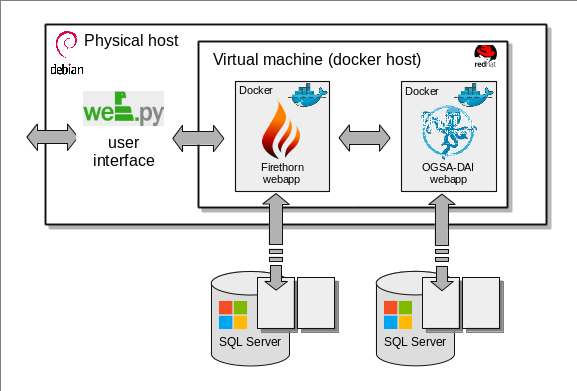}
\caption{First stage containerization (Tomcats but not Apache).
\label{fig:containerise1}}
\end{figure}

The first stage of containerization was to create Docker containers for the two
Tomcat web services, leaving the user interface web.py service running in the
Apache web server on the physical host.
The process of building the two Tomcat web service containers was automated using
the Maven Docker plugin\footnote{\burl{https://github.com/alexec/docker-maven-plugin}}. Figure~\ref{fig:containerise1}
illustrates this first stage.

\subsection{Using pre-packaged or in--house base images}

We ended up creating our own containers as the base images for our Tomcat
web services, rather than using the official
Java\footnote{\burl{https://hub.docker.com/_/java/}} and Tomcat\footnote{\burl{https://hub.docker.com/_/tomcat/}}
images available on the Docker registry.
This was a result of our early experiments with Docker
where we explored different methods of creating containers from simple Linux
base images and learned that creating our own base images gave
us much more control over the contents of our containers. The flexibility of the
container build system means that we are able to swap between base containers
by changing one line in a Docker buildfile and re-building.
This enabled us to test our containers using a variety of different base images,
and work towards standardizing on a common version of Python, Java and Tomcat
for all of our components.

Based on our experience, we would recommend that other projects follow a similar
route and define their own set of base images to build their containers, rather
than using the pre-packaged images available from the Docker registry. The
latter are ideal for rapid
prototyping, but there are some issues that mean they may not be suitable for
use in a production environment. Although the Docker project is working to
improve and to verify the official 
images\footnote{\burl{https://docs.docker.com/docker-hub/official_repos/}}, there is
still a lot of work to be done in this area.
The main issue with using a pre-packaged base image is that the contents of
containers are directly dependent on how the third party image was built and
what it contains.
Unless full details of what the third party image contains are available it can be
difficult to asses the impact of a security issue in a common component such as
OpenSSL\footnote{\burl{http://heartbleed.com/}}\textsuperscript{,}\footnote{\burl{https://cve.mitre.org/cgi-bin/cvename.cgi?name=cve-2014-0160}}
or glibc\footnote{\burl{https://www.kb.cert.org/vuls/id/457759}}\textsuperscript{,}\footnote{\burl{https://cve.mitre.org/cgi-bin/cvename.cgi?name=CVE-2015-7547}}\textsuperscript{,}\footnote{\burl{http://arstechnica.co.uk/security/2016/02/extremely-severe-bug-leaves-dizzying-number-of-apps-and-devices-vulnerable/}}
has on a system that is based on an opaque third party binary image.

\subsection{Ambassador pattern}

At this point in the project we also began to use the Docker ambassador 
pattern\footnote{\burl{http://docs.docker.com/engine/articles/ambassador_pattern_linking/}}
for managing the connections between our webapps and databases. The idea behind
the ambassador pattern is to use a small lightweight container running a simple
proxy service like
socat\footnote{\burl{http://www.dest-unreach.org/socat/}} to manage a
connection between a Docker container and an external service.

\begin{figure}[H]
\includegraphics[width=\columnwidth,clip=true,trim=5 0 0 5]{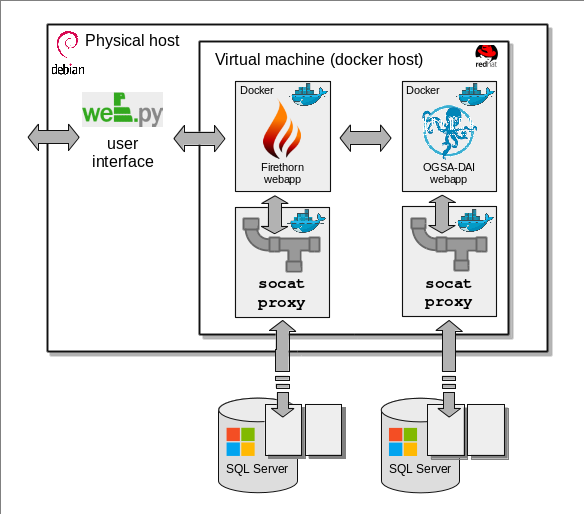}
\caption{Socat ambassadors for database connections.
\label{fig:socat}}
\end{figure}

In our case, the two socat proxies in Docker containers makes the relational
database appear to be running in another container on the same Docker host,
rather than on a separate physical machine. This enables our service
orchestration scripts to connect our web services to our database server using
Docker container links. The arrangement is shown schematically in 
Figure~\ref{fig:socat}.

At first glance, adding proxies like this may seem to be adding unnecessary
complication and increasing network latency for little obvious gain.
The benefit comes when we want to modify the system to support developers working
remotely on platforms outside the Institute network firewall, who need to be
able to run the set of services on their local system but still be able to connect to the
relational database located inside the firewall.

\begin{figure}[H]
\includegraphics[width=\columnwidth,clip=true,trim=5 0 0 5]{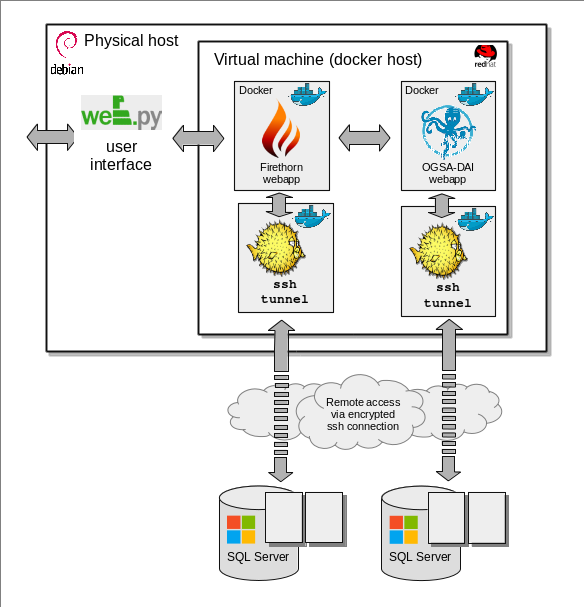}
\caption{SSH ambassadors for database connections.
\label{fig:sshambasadors}}
\end{figure}

In this scenario (illustrated schematically in Figure~\ref{fig:sshambasadors})
the \texttt{sql-proxy} containers are replaced by \texttt{sql-tunnel} containers
that use a tunneled ssh connection to link to the remote database located inside
the Institute network firewall.

\newtext{
The shell script for using the simple socat \texttt{sql-proxy} containers
creates a named instance of the \texttt{sql-proxy} container for each of the
database connections.
In the following example, we create two database proxy containers, one for the
metadata database and one for the userdata database. Each \texttt{sql-proxy}
container runs an instance of \texttt{socat} that listens on port 1433 on the
container and connects to port 1433 on the target database host:
}

\begin{singlespacing}
\begin{alltt}
docker run \textbackslash
  --detach \textbackslash
  --name metadata \textbackslash
  --env  targethost=\${datahost} \textbackslash
  firethorn/sql-proxy

docker run \textbackslash
  --detach \textbackslash
  --name userdata \textbackslash
  --env  targethost=\${datahost} \textbackslash
  firethorn/sql-proxy
\end{alltt}
\end{singlespacing}

\newtext{
Within the virtual network created by Docker, containers are accessible using
their names. So the configuration file for the Java web services can
use the names of the \texttt{sql-proxy} containers in the JDBC connection
url for the databases:
}

\begin{singlespacing}
\begin{alltt}
  jdbc:jtds:sqlserver://\textbf{metadata}/dbname
\end{alltt}
\end{singlespacing}
and
\begin{singlespacing}
\begin{alltt}
  jdbc:jtds:sqlserver://\textbf{userdata}/dbname
\end{alltt}
\end{singlespacing}

\newtext{
As far as the Java web services are concerned, they are making JDBC connections
to two machines on the local network called \texttt{metadata} and
\texttt{userdata}.
}

\newtext{
To re-configure the system to use remote tunneled connections to access the
databases, the deployment script can be modified to use instances of the
\texttt{sql-tunnel} container, passing in environment variables for the ssh user
name and host name used to create the ssh tunnel, and a volume mount of the
\texttt{SSH\_AUTH\_SOCK} Unix socket to allow the ssh client to use agent
forwarding\footnote{\burl{http://www.unixwiz.net/techtips/ssh-agent-forwarding.html\#fwd}}
for authentication:
}

\begin{singlespacing}
\begin{alltt}
docker run \textbackslash
  --detach \textbackslash
  --name metadata \textbackslash
  --env  tunneluser=\${tunneluser} \textbackslash
  --env  tunnelhost=\${tunnelhost} \textbackslash
  --env  targethost=\${datahost}   \textbackslash
  --volume \${SSH\_AUTH\_SOCK}:
    /tmp/ssh_auth_sock \textbackslash
  firethorn/sql-tunnel

docker run \textbackslash
  --detach \textbackslash
  --name userdata \textbackslash
  --env  tunneluser=\${tunneluser} \textbackslash
  --env  tunnelhost=\${tunnelhost} \textbackslash
  --env  targethost=\${datahost}   \textbackslash
  --volume \${SSH_AUTH_SOCK}:
    /tmp/ssh_auth_sock \textbackslash
  firethorn/sql-tunnel
\end{alltt}
\end{singlespacing}

\newtext{
Each \texttt{sql-tunnel} container runs an instance of the \texttt{ssh}
client that listens on port 1433 on the container and creates an encrypted
tunneled connection via the ssh gateway host to port 1433 on the target database
host.
}

\newtext{
Because the \texttt{sql-tunnel} containers function as drop-in replacements for
the \texttt{sql-proxy} containers, as far as the rest of the system is
concerned, nothing has changed. The configuration files use the same URLs for
the JDBC database connections, and as far as the Java web services are concerned,
they are still making JDBC connections to two machines on the local
network called \texttt{metadata} and \texttt{userdata}.
}

\newtext{
Obviously, using tunneled ssh connections for database access adds significant
latency to the system, and would not be appropriate for a production system.
However, based on our experience, using tunneled ssh connections for database
access works well for development and testing.
}

\subsection{Python GUI and Python testing}

The final stage in the migration to Docker containers was to wrap the web.py
user interface service in a container and add that to our set of images.

\begin{figure}[H]
\includegraphics[width=\columnwidth,clip=true,trim=5 0 0 5]{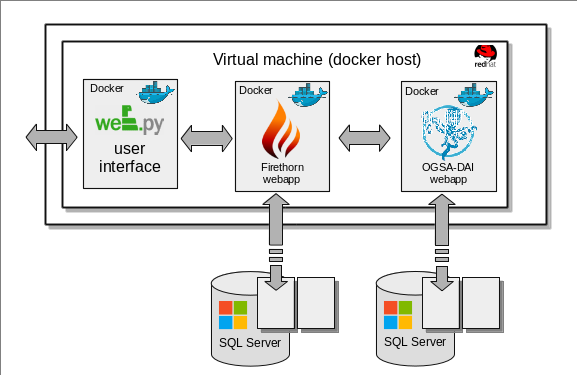}
\caption{Adding the web.py container.\label{fig:webpy}}
\end{figure}

The web.py web user interface container is built starting with a basic Ubuntu
image and building on that with a series of containers that add the Apache
webserver, the Python language, a set of Python libraries and finally the web.py
web application itself.

An additional web.py web interface was later developed for a separate
project (Gaia European Network for Improved User Services;
e.g.~\citealt{2016arXiv160307347H}), which usedthe distributed querying feature of Firethorn. Because of the
separation of the interfaces, Firethorn web services and databases into
containers and the modular design of Docker systems, attaching this new
interface container to the existing set of Docker containers was seamless.
Linking a configuration file and startup script when running the web application
-- a common technique when deploying web application containers which makes the
interchange of components in the system chain easier -- was also used in both.

Another example of a top level container used in our system was the testing
suite that we used to test our system for performance and accuracy, also written
in Python. This consisted of a number of possible tests, which would each launch
an instance of the core web service containers, as well as a number
of other containers required for the tests, for databases to log results,
or for loading and running the test suite code. By the end of the project we employed a set
of bash scripts that allowed us to run a one line command to start the
required test, which we would run on any virtual machine.
These were long running tests, which helped
us gauge how a system using Docker containers would behave and scale with
large data volumes and long term up-time and whether Docker as a technology was
production--ready or not. The full test deployment also included a local MySQL
database deployed in a container alongside the Python test application for
storing test results.

\begin{figure}[H]
\includegraphics[width=\columnwidth,clip=true,trim=5 0 0 5]{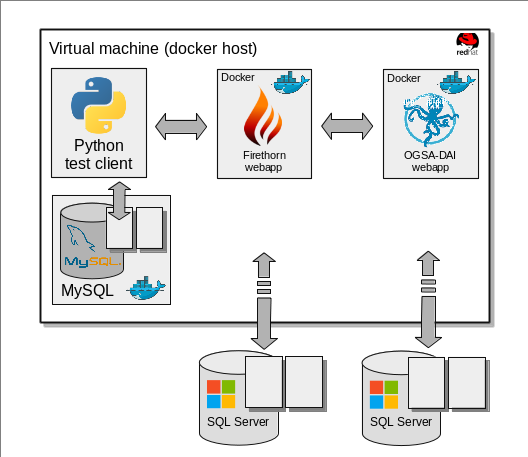}
\caption{Python test suite configuration.\label{fig:pyrothorn}}
\end{figure}

The result is a set of plug--and--play containers for each component in our system
that can be swapped and replaced with different versions or different
implementations by modifying the scripts that manage the container orchestration. 


A live deployment would include the web.py web application for the user
interface, and use socat proxies to connect to the local relational databases.
In the test and development scenarios we replace the web.py web application with
a Python test client connected to a local MySQL database running in a container,
and in some cases we also replaced the connection to the SQL Server metadata
database for our Firethorn web service with a PostgreSQL database running in a
container.

\subsection{Orchestrating build and deployment}

All of our containers are managed by a set of shell scripts which are
included and maintained as part of the project source code. The Docker build
scripts and the container orchestration scripts required to build and deploy a
full set of services for each of the use cases are all stored in our source
control repository alongside the source code for the rest of our project.
Automating the service deployment, and treating the build and deployment scripts
as part of the core project source code is a key step towards implementing what
is referred to as \emph{Programmable Infrastructure} or \emph{Infrastructure
as code}\footnote{\burl{http://devops.com/2014/05/05/meet-infrastructure-code/}}\textsuperscript{,}\footnote{\burl{https://www.thoughtworks.com/insights/blog/infrastructure-code-reason-smile/}}.

\newtext{
The bash scripts described in previous sections are used to deploy and link
Docker container instances to create the required configuration of containers and
services. We have recently started to experiment with Docker
Compose\footnote{\burl{https://docs.docker.com/compose/}} which makes this
process much simpler and clearer. 
}

\newtext{
Compose allows you to define a set of container configurations in a
\texttt{YAML}\footnote{\burl{http://www.yaml.org/}} file, where all the options
that were defined in the shell scripts and passed as parameters to the Docker
\texttt{run} command, are now defined in the \texttt{YAML} configuration file.
Which means a complete set of inter-linked containers can be initialised with a
single \texttt{docker-compose} command:
}

\begin{singlespacing}
\begin{alltt}
    docker-compose run <service>
\end{alltt}
\end{singlespacing}

\newtext{
Where \texttt{<service>} refers to one of the container instances defined in the
\texttt{YAML} configuration file. Compose simplifies the process of
configuring, initialising and linking containers, and overall the process of building
development, testing, and staging environments as well as continuous integration
workflows.
}

\newtext{
However, at the time of writing we have only just started experimenting with
Compose and we do not have enough experience with it to provide a more thorough
description of its use.
}

\subsection{Cross platform deployment}  

One of the key reasons for choosing Docker to deploy our systems was to be able
to deploy the software reliably and repeatably on a range of different
platforms.

In our project the software has to be able to run on a number of
different platforms, including the developer's desktop computer, the integration
test systems and at least two different live deployment environments.
A key requirement of our project is that the software must be able
to be deployed at a number of different third party data centres, each of which
would have a slightly different operating system and runtime environment. 

If we relied on manual configuration for the target platform and runtime
environment, then over time it is inevitable that they will end up being
slightly different. Even something as simple as the version of Java or Tomcat
used to run the web application can be difficult to control fully. We could, in
theory, mandate a specific version and configuration of the software stack used
to develop, test and deploy our software. In reality, unless the platform is
created and managed by an automated process, then some level of discrepancy will
creep in, often when it is least expected.

There are a number of different ways of achieving this level of automation.
A common method of managing a large set of systems is to use an automated
configuration management tool, such as Puppet\footnote{\burl{https://puppetlabs.com/}} or 
Chef\footnote{\burl{https://www.chef.io/chef/}}, to manage the system configuration
based on information held in a centrally controlled template. Another common
practice is to use a continuous integration platform such as 
Jenkins\footnote{\burl{https://wiki.jenkins-ci.org/}} to automate the testing and
deployment. These techniques are not mutually exclusive, and it is not unusual to use an
automated configuration management tool such as Puppet to manage the (physical
or virtual) hardware platform, in combination with a continuous integration
platform such as Jenkins to manage the integration testing, and in some cases
the live service deployment as well. However, these techniques are only really
applicable when one has direct control over the deployment platform and the
environment around it.
In our case, we knew that although we had control over the environment for
our own deployments, we would not have the same level of control over
deployments at third party sites.

\section{Issues found and lessons learned}
\label{sec:issues}

It is of course expected that issues and problems arise 
when using new technologies for the first time.
These might be caused by mistakes made while climbing the learning curve or by
software bugs in the technology itself, which may have not been uncovered yet
while adoption of the technology is still growing, and all possible usages of it
have not been visited yet.
We document here an example of one of the issues we encountered, including how
we solved it.

\subsection{Memory issue}

As part of our Firethorn project we developed a testing suite written in Python
as mentioned above. This suite included some long--running tests, which iterated
a list of user submitted SQL queries that had been run through our systems in
the past, running the same query via a direct route to the database as well as
through the new Firethorn system and comparing the results. This list scaled up
to several thousand queries, which meant that a single test pass for a given
catalogue could take several days to complete. The issue we encountered here was
that the Docker process was being killed after a number of hours, with `Out of
memory' error messages. An initial attempt at solving the problem was to set
memory limits on all of our containers, which changed the symptoms and then
caused our main Tomcat container to fail with memory error messages.
After a few iterations of attempting to run the chain with different
configurations, the solution was found through community forums, when we
discovered that several other people were encountering the same symptoms with
similar setups. Specifically, the problem was due to a memory leak, caused by
the logging setup in the version of Docker that we were using (1.6). Output sent
to the system \texttt{stdout} was being stored in memory causing a continuous
buffer growth resulting in a memory
leak\footnote{\burl{https://github.com/docker/docker/issues/9139}}\textsuperscript{,}\footnote{\burl{https://github.com/coreos/bugs/issues/908}}.

The solution that we adopted was to use the \texttt{volume} option to
send the system output and logs from our container processes to a directory
outside the container: 

\begin{alltt}
docker run
  ...
  --volume "/var/logs/firethorn/
      :/var/local/tomcat/logs" 
  ...
  "firethorn/firethorn:2.0"
\end{alltt}

We learned several valuable lessons through the process of researching how other
developers managed these problems, for example, the approach to logging where
the logs of a container are stored separately from the container itself, making
it easier to debug and follow the system logs.  In addition, we benefited from
learning how and why limiting memory for each container was an important step
when building each of our containers.

A fix for this issue was added to the Docker source code in November 2015\footnote{\burl{https://github.com/docker/docker/pull/17877}}
and released in Docker version 1.10.

In addition, Docker added a pluggable driver based framework for handling
logging\footnote{\burl{https://github.com/docker/docker/pull/10568}}\textsuperscript{,}\footnote{\burl{https://blog.logentries.com/2015/06/the-state-of-logging-on-docker-whats-new-with-1-7/}}
which provides much more control over how logging output from processes
running in the container is handled\footnote{\burl{https://docs.docker.com/engine/reference/logging/overview/}}.

\subsection{Docker community}

More important than an analysis of the issues themselves is the understanding of
the process undertaken to discover and solve them. 
An important point to make here, is in regard to the open source nature and
culture of Docker and the Docker community.
The main takeaway from this was finding how to go about solving issues
related to containers and figuring out how the preferred method of implementing
a certain feature is easy enough as doing a search of the keywords related to
what you need.  This can be done by either using a generic search engine or
visiting the sources where the main Docker community interaction takes
place\footnote{\burl{https://forums.docker.com/c/general-discussions/general}}\textsuperscript{,}\footnote{\burl{http://stackoverflow.com/questions/tagged/docker}}\textsuperscript{,}\footnote{\burl{https://github.com/docker/docker/issues}}.

Like many open source solutions, Docker has an active open source
community behind it which enables users to find and fix issues more efficiently.
An \newtext{active} open source community means it is more likely that any issue
you might find has already been encountered by someone else, and just as likely that it has
been solved officially (as part of a bug fix in Docker release) or unofficially
(by some community member describing how they solved the problem).
\newtext{The memory issue described in the previous section is an example of
how using community resources helped us to find how others who had encountered
the same problem and how they had solved it.}

While Docker's source code is open to the public, perhaps more importantly so is
its issue tracking system. Apart from the fact that issues will get raised and
solved quicker naturally with more eyes on them, another advantage for the users
of such a platform is that they get the opportunity to contribute and help steer
the direction it takes, by either raising issues or adding comments to the issue
tracking system or the discussion forums. This leads to the targets for each new
release being closely tied with what the majority of the community raises as
important issues or requests for future enhancements.

Another key point to note is how we benefited from Docker’s support team as well
as the number of early adopters. We decided to take up Docker at an early stage,
which can be considered its `bleeding-edge' phase (version 1.6), at which
point it was more likely to discover issues. However, with the large team and
strong technological support of its developers, as well as the significant
number of early adopters, new releases to solve bugs or enhance usability and
performance were issued frequently. Consequently, after some research, we
realized that many of the issues we found, whether they could be considered bugs
or usability improvements needed, were often fixed in subsequent releases,
meaning that by updating our Docker version they would be solved.

\newtext{
Active participation in the community by the project developers and the
fostering of an `inclusive' atmosphere where users feel confident to submit
bugs, request changes and post comments all contributes to the success of the
project. This is true for many, but by no means all, open source projects, Linux
itself being a prime example. Just making the source code accessible does not
in itself guarantee the successful adoption of a project. In our experience,
the more active and responsive the core project developers are to input
from users, the more likely it is that a project will be successful and be
widely adopted. This has certainly been the case so far with the Docker
project.
}

\subsection{Learning curve}

\newtext{
Getting started with creating basic containers was relatively easy, starting
with the simple images available from Docker Hub, along with the
extensive documentation and user guides, as well as the community forums.
}

\newtext{
In the process of creating our containers we started with base images for the
applications we wanted to create, for example using the official Tomcat image,
looking at the source code for the \texttt{Dockerfile}, figuring out how they
were put together and then developing our own version once we understood how
they worked.
}

\newtext{
Understanding concepts like the isolation of each process of an application, how
to link containers and expose ports, as well as how best to handle logging and
resource usage, develop later as a result of using Docker containers
for different applications and exloring the comments and advice available on
the Docker community sites and third party blogs.
}

\section{Conclusion}
\label{sec:conclusion}

As mentioned throughout this paper, some of the main takeaways we noted from the
use of Docker in development and production are the ease it provides in bundling
components together, promoting re-usability, maintainability and faster
continuous integration environments.
\newtext{
We also noted how Docker improved collaboration between developers, specifically
by providing a standard testing and deployment environment. Sharing code which
is then compiled and executed on different environments has the potential to
behave differently, while even the setup of such an environment can be
cumbersome. By using Docker containers, developers need only share a Docker
image or Dockerfile, which guarantees the environment will be the same.
}

In addition, \newtext{based on our own experience of working with Docker and
from talking about Docker with colleagues on other projects} the openness of
Docker and its community has contributed to its popularity in both science and
business systems.

Based on our experience in development and production for the Firethorn and
IVOA\TeX\ projects, we anticipate a rapid growth of interest and usage of Docker
and container-based solutions in general. We expect that this will be the case
for both developing and deploying systems as a replacement or complementary to
exiting hardware virtualization technologies, in enabling reproducible
science and in systems that allow scientists to submit their own code to data
centres. Docker can potentially help with this, as it
provides the tools and simplicity that scientists need to recreate the
environment that was used to generate a set of test results.

In terms of the future of Docker in relation to the OCI, there is the potential
for a common container standard to emerge, with the Docker project playing a
leading role in the shaping of this standard.
It should be noted that as explicitly stated by the OCI, given the broad
adoption of Docker, the new standard will be as backward compatible as possible
with the existing container format. Docker has already been pivotal in the
OCI by donating draft specifications and code, so we expect any standard that
emerges from this process will be closely tied with what exists now in Docker.
 
Docker is not the perfect solution, and scientists or system engineers must
decide when and if it is a suitable tool for their specific needs. It is most
applicable in situations where reproducibility and cross-platform deployment are
high on the list of requirements.

When deciding on whether to adopt a container technology such as Docker our
experience would suggest that the benefits in terms of re-usability,
maintainability and portability represent a significant benefit to the project
as a whole and in most cases we would expect the benefits to outweigh the costs
in terms of learning and adopting a new technology. 

\section*{Acknowledgments}
\label{sec:acknowledgments}
 
The research leading to these results has received funding from: (i)
 The European Community's Seventh Framework Programme (FP7-SPACE-2013-1)
  under grant agreement n°606740; (ii) The European Commission Framework Programme Horizon 2020 Research and
  Innovation Action under grant agreement n. 653477; and (iii)
The UK Science and Technology Facilities Council under grant numbers
  ST/M001989/1, ST/M007812/1, and ST/N005813/1

The authors would like to thank the reviewer for a comprehensive and detailed
list of revision recomendations.

\section*{References}
\label{sec:references}

\bibliography{mybibfile}

\end{document}